\documentclass{aa}

  \newcommand\dodouble[1]{ } 

\dodouble{\documentclass[referee]{aa}}

\usepackage{natb2e}

\renewcommand\citep[1]{(\citealt{#1})}

\newcommand\zzz[3]{#2}  

\zzz{\usepackage{epsfcopy}}{\usepackage{epsfcopy}}{}

\newcommand\submitstyle[2]{#2} 

\submitstyle{\newcommand\noarXiv[1]{}}{\newcommand\noarXiv[1]{#1}}

\newcommand\centreline{\centerline}
\newcommand\hMpc{\mbox{$\,h^{-1}$ Mpc}}

\newcommand\hGpc{\mbox{$h^{-1}$ Gpc}}

\newcommand\Omm{\Omega_{\mbox{\rm \small m}}}
\newcommand\ltapprox{\,\lower.6ex\hbox{$\buildrel <\over \sim$} \, }

\newcommand\llss{L}

\newcommand\Pzero{P_{\mbox{\rm \small o-free}}}
\newcommand\Pllss{P_{\mbox{\rm \small L}}}

\newcommand\dpm{d_{\mbox{\rm \small pm}}}
\newcommand\dperp{d_\perp}
\def\e{ {\scriptstyle \times} 10^}

\newcommand\frtoday{Le\space\number\day\space\ifcase\month\or
  janvier\or f\'evrier\or mars\or avril\or mai\or juin\or
  juillet\or ao\^ut\or septembre\or octobre\or novembre\or 
  d\'ecembre\fi\space \number\year}

\submitstyle{\newpage}{}
\newcommand\treject{
\begin{table*}
\caption[]{ 
Minimum rejection values 
$1-\Pzero\Pllss$ of the bold curves in 
Figs~\protect\ref{f-rejecten} and \protect\ref{f-rejeconep}.
\label{t-reject}
}
\begin{tabular}{ccccc cc}
\hline
{}{$\protect\Omm$} & 
{}{$\Omega_\Lambda$} & 
{}{$\Omega_\kappa$} & {}{$z$} & 
{}{$\llss$}& 
{}{$\Delta\protect\llss $}& 
{}{$\min\{1-\protect\Pzero \protect\Pllss \}$} \\
{}{} & {}{} & {}{} & 
{}{(\protect{\hMpc})}& 
{}{(\protect{\hMpc})} 
 \\
0.2 &0.7 &-0.1 &3 & 130 &10  & 31\% \\
0.4 &0.5 &-0.1 &3& 130 &10  &  20\% \\
0.4 &0.7&+0.1  &3& 130 &10  &  24\% \\
0.4 &0.5 &-0.1 &2& 130 &10  &  14\% \\
0.2 &0.7&-0.1 &3 & 130 &1.3  & 99.7 \% \\ 
0.4 &0.5 &-0.1 &3& 130 &1.3  & 94\% \\
0.4 &0.7&+0.1  &3& 130 &1.3  & 97\% \\
0.4 &0.5 &-0.1 &2& 130 &1.3  & 80\% \\
\hline
 \end{tabular} 
\end{table*}
\submitstyle{\clearpage}{}
}  

\submitstyle{\newpage}{}
\newcommand\trejecttwo{
\begin{table*}
{
\caption[]{ 
Minimum rejection values $1-\Pzero\Pllss$, 
as for Table~\protect\ref{t-reject}, but for a larger range in 
$\Omm, \Omega_\Lambda$, with $z=3,$ 
$\llss=130${\hMpc}, $\Delta \llss=1.3${\hMpc} fixed.
\label{t-rejecttwo}
}
\begin{tabular}{cccc cccc c}
\hline
{}{$\protect\Omm$} & 
{}{$\Omega_\Lambda$} & 
{}{$\Omega_\kappa$} & 
{}{$\min\{1-\protect\Pzero \protect\Pllss \}$} &
{}{$\protect\Omm$} & 
{}{$\Omega_\Lambda$} & 
{}{$\Omega_\kappa$} & 
{}{$\min\{1-\protect\Pzero \protect\Pllss \}$} \\
\multicolumn{8}{c}{\mbox{dependence on $\Omm$}} \\
    0.1 &   0.8 &  -0.1 & 100.0\% &   0.1 &   1.0 &   0.1 & 100.0\% &\\
    0.2 &   0.7 &  -0.1 &  99.7\% &   0.2 &   0.9 &   0.1 &  99.9\% &\\
    0.3 &   0.6 &  -0.1 &  97.6\% &   0.3 &   0.8 &   0.1 &  99.2\% &\\
    0.4 &   0.5 &  -0.1 &  93.9\% &   0.4 &   0.7 &   0.1 &  97.2\% &\\
    0.5 &   0.4 &  -0.1 &  89.9\% &   0.5 &   0.6 &   0.1 &  93.8\% &\\
    0.6 &   0.3 &  -0.1 &  86.4\% &   0.6 &   0.5 &   0.1 &  89.5\% &\\
    0.7 &   0.2 &  -0.1 &  83.4\% &   0.7 &   0.4 &   0.1 &  84.9\% &\\
    0.8 &   0.1 &  -0.1 &  77.5\% &   0.8 &   0.3 &   0.1 &  80.1\% &\\
    0.9 &   0.0 &  -0.1 &  70.5\% &   0.9 &   0.2 &   0.1 &  75.3\% &\\
\multicolumn{8}{c}{\mbox{a higher value of $|\Omega_\kappa|$}} \\
    0.1 &   0.7 &  -0.2 & 100.0\% &   0.1 &   1.1 &   0.2 & 100.0\% &\\
    0.2 &   0.6 &  -0.2 & 100.0\% &   0.2 &   1.0 &   0.2 & 100.0\% &\\
    0.3 &   0.5 &  -0.2 & 100.0\% &   0.3 &   0.9 &   0.2 & 100.0\% &\\
    0.4 &   0.4 &  -0.2 &  99.9\% &   0.4 &   0.8 &   0.2 & 100.0\% &\\
    0.5 &   0.3 &  -0.2 &  99.7\% &   0.5 &   0.7 &   0.2 & 100.0\% &\\
    0.6 &   0.2 &  -0.2 &  99.2\% &   0.6 &   0.6 &   0.2 &  99.9\% &\\
    0.7 &   0.1 &  -0.2 &  98.3\% &   0.7 &   0.5 &   0.2 &  99.6\% &\\
    0.8 &   0.0 &  -0.2 &  97.2\% &   0.8 &   0.4 &   0.2 &  99.0\% &\\
    0.9 &  -0.1 &  -0.2 &  95.6\% &   0.9 &   0.3 &   0.2 &  98.0\% &\\
\hline
 \end{tabular} 
} 
\end{table*}
}  

\newcommand\tredacc{
\begin{table}
\protect{
\caption[]{Statistical redshift accuracy 
$\Delta z$
needed for positions
of low redshift structures used for calibrating $\llss$.
For a given redshift $z$ and given metric parameters, 
the uncertainty in redshift $\Delta z$ corresponding
to $\Delta \llss=1.3${\hMpc} (1\% precision in $\llss=130${\hMpc})
is given. This shows that over the domain shown, a precision of
$\Delta z \ltapprox 5\e{-4}$ 
in redshift is required. 
\label{t-redacc} 
}
$$\begin{array}{cccc cccc} 
\hline
\Omm & \Omega_\Lambda & \multicolumn{3}{c}{z} \\
        &         &    0.1 &       0.3 &       0.5 &   \\ 
   0.3 &    0.0 &    0.0005 &    0.0006 &    0.0007 &   \\ 
   0.3 &    0.7 &    0.0005 &    0.0005 &    0.0006 &   \\ 
   1.0 &    0.0 &    0.0005 &    0.0006 &    0.0008 &   \\ 

\hline
\end{array}$$ 
} 
\end{table} 
}  

\submitstyle{\newpage}{}
\newcommand\fdeltad{
\begin{figure}
{\centering 
 \centreline{
\zzz{\epsfxsize=7cm\epsfbox[29 59 462 499]{"`gunzip -c 
M10394f1.eps.gz"} }
{\epsfxsize=7cm\epsfbox[29 59 462 499]{"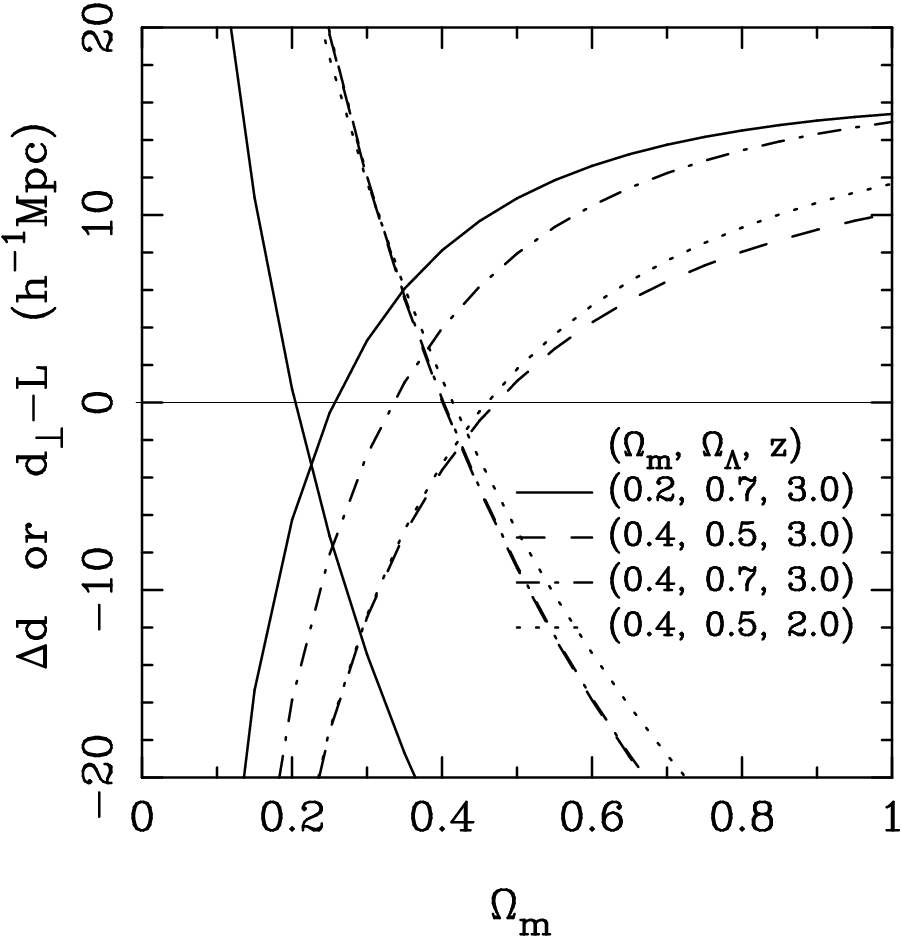"}}{}  }
\caption[M10394f1.eps]{
Differences between tangential and radial lengths implied
by assuming a flat universe, 
$\Delta d(z,\Omm,1-\Omm)$ (eq.~\protect\ref{e-diff}), and
differences between the tangential ruler size and its known size,
$\dperp(z,\delta\theta,\Omm,1-\Omm) - \llss$, both shown in
comoving {\hMpc} as a function of $\Omm$. The $\Delta d$ curves
increase with $\Omm$; the $\dperp - \llss$ curves 
decrease with $\Omm.$ Various input hypotheses
for hyperbolic and spherical universe models and redshifts $z$ 
are shown with different curves as labelled.  The points where the 
increasing curves pass through zero are those where a flat universe
implies independence of the ruler size from orientation. 
The points where the decreasing curves pass through zero are those
where the ruler size is the known (e.g. zero redshift) 
size of the ruler.
\label{f-deltad}
}}
\end{figure}
} 

\submitstyle{\newpage}{}

\newcommand\frejecten{
\begin{figure}
{
\centering 
\centreline{
\zzz{\epsfxsize=7cm\epsfbox[ 36 49 462 491]{"`gunzip -c 
M10394f2.eps.gz"} }
{\epsfxsize=7cm\epsfbox[ 36 49 462 491]{"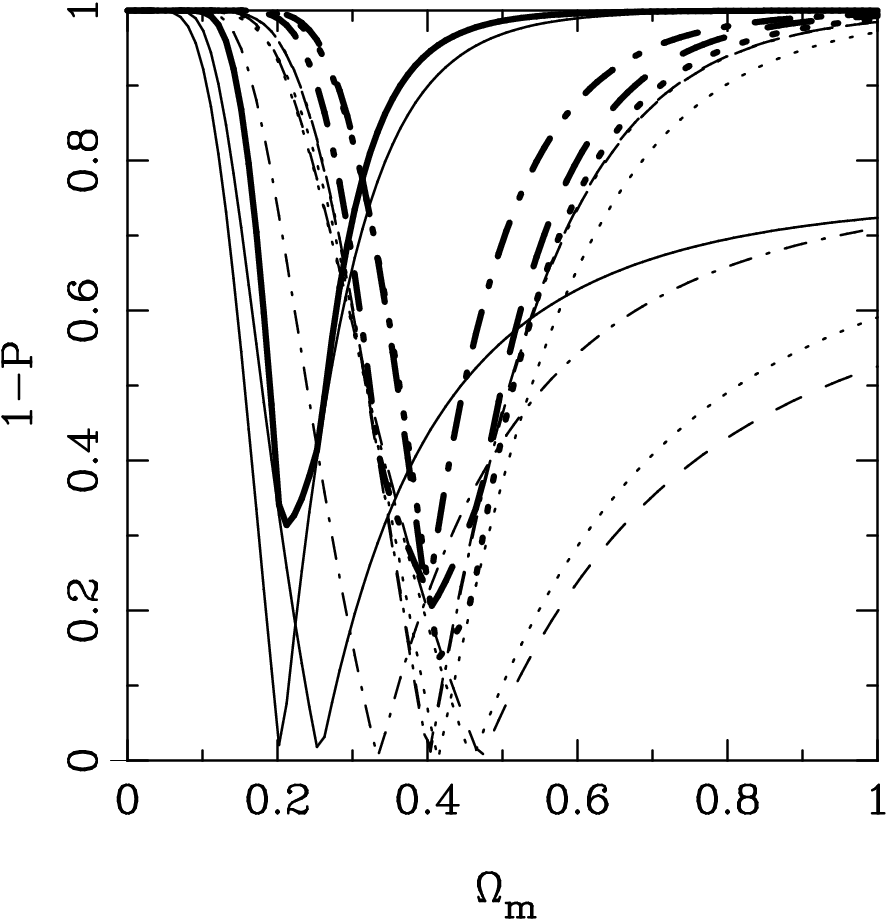"}}{}  }
\caption[M10394f2.eps]{Rejection probabilities based on the results 
in Fig.~\protect\ref{f-deltad}, assuming that 
$\llss=130${\hMpc}, $\Delta\llss=10${\hMpc}, and using 
equations (\protect\ref{e-pzero}), (\protect\ref{e-pllss}) and
(\protect\ref{e-pboth}). Curve styles are as for 
Fig.~\protect\ref{f-deltad}. The thin curves are for 
$1-\Pzero$ and $1-\Pllss$, the bold curves are for 
$1-\Pzero\Pllss$.
\label{f-rejecten}
}}
\end{figure}
} 

\submitstyle{\newpage}{}
\newcommand\frejeconep{
\begin{figure}
{\centering 
 \centreline{
\zzz{\epsfxsize=7cm\epsfbox[ 36 49 462 491]{"`gunzip -c 
M10394f3.eps.gz"} }
{\epsfxsize=7cm\epsfbox[ 36 49 462 491]{"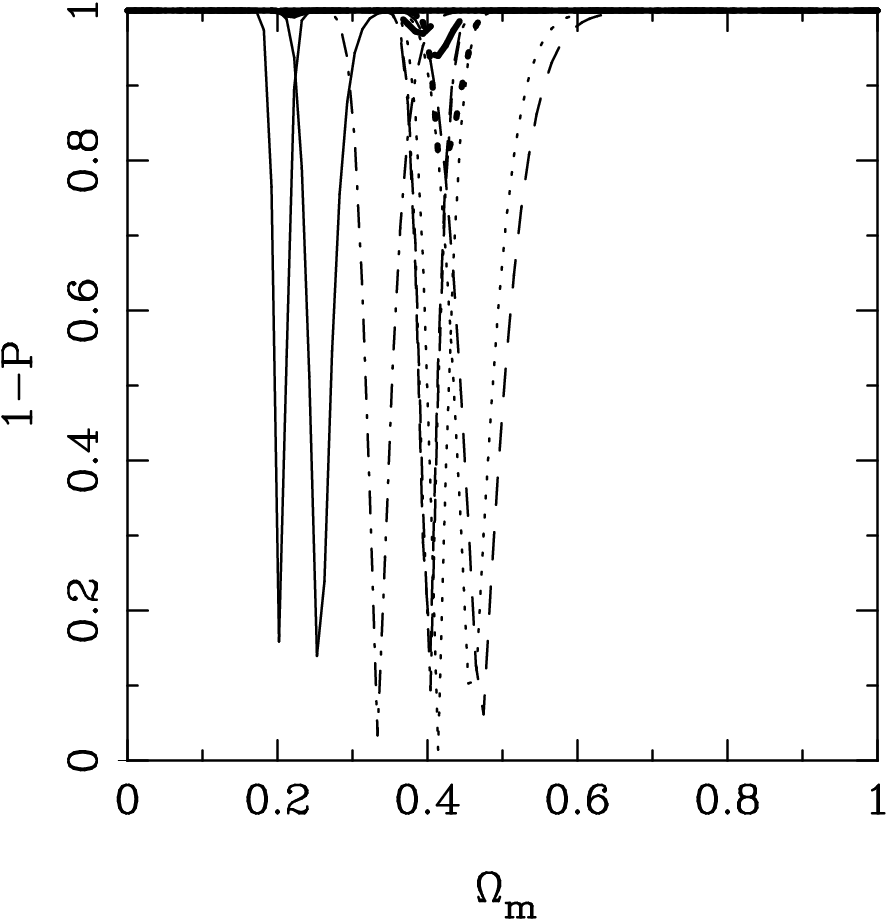"}}{}  }
\caption[M10394f3.eps]{As for Fig.~\protect\ref{f-rejecten}, 
for $\llss=130${\hMpc}, $\Delta\llss=1.3${\hMpc} (i.e. 1\% precision).
The $1-\Pzero\Pllss$ curves in bold are barely visible, close to 
the 100\% rejection limit. Their minima are listed in 
Table~\protect\ref{t-reject}.
\label{f-rejeconep}
}}
\end{figure}
} 
\submitstyle{\clearpage}{}

\begin{document}

\title{How to distinguish a nearly flat
Universe from a flat Universe using the 
orientation independence of a comoving standard ruler}

\author{Boudewijn F. Roukema\inst{1,2}}

\institute{Inter-University Centre for 
Astronomy and Astrophysics 
{Post Bag 4, Ganeshkhind, Pune, 411 007, India}
\and
DARC, Observatoire de Paris-Meudon, 5, place Jules Janssen,
F-92195, Meudon Cedex, France
{\em (boud.roukema@obspm.fr)} }

\titlerunning{Using a standard ruler to disprove flatness}
\authorrunning{B.~F.~Roukema}

\date{\frtoday}

\abstract{
Several recent observations 
using standard rulers and standard candles now suggest, either 
individually or in combination, that the Universe is close to flat, 
i.e. that the curvature radius is about as large as the horizon 
radius ($\sim 10${\hGpc}) or larger. 
Here, a method of distinguishing an 
almost flat universe from a precisely flat universe using a single 
observational data set, without using any microwave background 
information, is presented. The method 
(i) assumes that a standard ruler should have no preferred orientation 
(radial versus tangential) to the observer, 
and (ii) requires that the (comoving) length of the standard ruler 
be known independently (e.g. from low redshift estimates). 
The claimed feature at fixed {\em comoving} length 
in the power spectrum of density perturbations, detected among 
quasars, Lyman break galaxies or other high redshift objects, 
would provide an adequate standard candle to prove that 
the Universe is curved, if indeed it is curved. For example, 
a combined intrinsic and 
measurement uncertainty of 
$1\%$ in the length of the standard ruler $\llss$ applied at 
a redshift of $z=3$ would 
distinguish an hyperbolic $(\Omm=0.2,\Omega_\Lambda=0.7)$ 
or a spherical $(\Omm=0.4,\Omega_\Lambda=0.7)$ 
universe from a flat 
one to $1-P > 95\%$ confidence. 
}

\maketitle

\section{Introduction}

In the Friedmann-Lema\^{\i}tre-Robertson-Walker model \citep{Wein72}, 
the Universe is an almost 
homogeneous 3-manifold of constant curvature.
This manifold may be the 3-hyperboloid $H^3$, flat Euclidean space $R^3$ 
or the 3-sphere $S^3$, or a quotient manifold of one of these, 
e.g. the 3-torus $T^3$ \citep{Schw00,Schw98}. 

A directly geometrical way to measure curvature is by
a standard candle or a standard ruler, i.e. a class of objects
of which the intrinsic brightness or comoving length scale is believed
to be fixed. 
Several recent applications of standard candles or standard rulers
include 
the standard candle applications of
\cite{SCP9812} and \cite{HzS98}, the standard ruler applications 
of \cite{RM00a,RM00b}, 
\cite{Boom00a} and \cite{Maxima00a}, which use geometrical information
in the tangential direction,
and the standard ruler application of 
\cite{BJ99}, which uses geometrical information in the
radial direction.

These recent measurements, individually or in combination, favour
an `almost' flat local universe. However, \cite{Jaffe00} reject a flat
universe to just under 95\% significance, 
finding $\Omega_{\mbox{\rm \small tot}} =
1.11^{+0.13}_{-0.12}$ (`95\% confidence'). 

Whether the \cite{Jaffe00} result is just due to random or 
systematic error and the observable Universe is in fact 
flat to high precision as predicted by many models of inflation,
or whether the Universe really is measurably curved, 
distinguishing an almost (but not) flat
universe from an `exactly' flat model will require considerably 
more precise techniques than have been previously applied.

Given the claimed existence of a comoving standard ruler 
\citep{BJ99,RM00a,RM00b}, what could possibly be the most model
free technique for testing the flatness hypothesis 
is presented here, for the
case where the Universe is curved and independence of orientation of
a comoving standard ruler is used to refute the flat universe hypothesis.
Since
\begin{list}{(\roman{enumi})}{\usecounter{enumi}}
\item a standard ruler should have no preferred orientation
(radial versus tangential) with respect to the observer, and
\item the radial and tangential comoving 
distances differ by a $\sin$ or
$\sinh$ factor if the Universe is curved [eq.~(\protect\ref{e-defpm})],
\end{list}
then if the lengths of 
a standard ruler in the tangential and 
radial directions are proved to be unequal under the assumption of 
a flat universe, it might be thought that this would falsify the flat 
universe hypothesis. However, as shown below, the additional constraint 
that 
\begin{list}{(\roman{enumi})}{\usecounter{enumi}}
\addtocounter{enumi}{2}
\item the standard ruler should have a known, fixed comoving value
\end{list}
is required in order to falsify the flat universe hypothesis.

Since the claimed standard ruler is in the linear regime of density
perturbations, and is presumably a primordial feature in the power
spectrum (if real), its inability to evolve in comoving length scale in less
than a Hubble time would make it free of the evolutionary effects
present for standard rulers or standard candles defined by collapsed
objects.

\cite{AP79}, \citet{Phil94}, \citet{MSuto96} and 
\citet*{Ball96} have previously pointed out the potential usefulness
of (i) and (ii), and have suggested applying these
at quasi-linear or non-linear scales, i.e. at $r \la 10${\hMpc}, including
some analysis of how to try to separate out peculiar velocity effects.
However, they did not discuss
how to lift the degeneracy in the two curvature 
parameters $(\Omm,\Omega_\Lambda)$ which remains after using (i) and (ii),
and the problem of evolution of the 
$r \la 10${\hMpc} auto-correlation functions of galaxies and quasars
offers potentially serious systematic uncertainties.

In contrast, by using a comoving standard ruler in the linear regime, 
the constancy in the comoving scale over a Hubble
time [(iii) above] provides an additional constraint in the 
$(\Omm,\Omega_\Lambda)$ plane, which is necessary 
in order to try to reject the flatness hypothesis. 
This is shown below (Fig.~\ref{f-deltad}, \S\ref{s-method}). 

Moreover, use of a linear regime comoving ruler also implies that
peculiar velocity effects become negligible.

Note that since the method uses data from a single survey, 
it is qualitatively quite different from the concept of 
cosmic complementarity \citep{EisHTeg98,Linew98}. 

The distance relations are reviewed in \S\ref{s-distances}, 
a method of illustrating the principle is explained in 
\S\ref{s-method}, results are presented in \S\ref{s-results},
and conclusions are made in \S\ref{s-conclu}.

\section{Distance relations} \label{s-distances}

Using the terminology of \cite{Wein72}, the distance of use
in the radial direction in comoving coordinates 
is the 
{\em proper distance} 
[eq.~(14.2.21), \citealt{Wein72}; 
denoted $R \chi$ by \cite{Peeb93}, eq.~(13.28)], 
\begin{equation}
d(z) 
= {c \over H_0} \int_{1/(1+z)}^1 
{ \mbox{\rm d}a \over a 
\sqrt{\Omm /a - \Omega_\kappa + \Omega_\Lambda a^2} },
\label{e-defdprop}
\end{equation}
where $c$ is the speed of light, $H_0$ is the Hubble constant, 
$\Omm$ is the matter density divided by the critical density,
$\Omega_\Lambda$ is the cosmological constant, 
$z$ is redshift, and the dimensionless curvature is
\begin{equation}
\Omega_\kappa \equiv \Omm + \Omega_\Lambda -1.
\end{equation}

The curvature radius is then 
\begin{equation}
R_C \equiv {c \over H_0} { 1 \over \sqrt{ | \Omega_\kappa| } }.
\end{equation}

For likely values of the
curvature parameters, i.e. 
($\Omm\approx 0.3, \Omega_\Lambda \approx 0.7$), the
value of $R_C$ is constrained to 
$
R_C \ga 10${\hGpc}
for a curvature 
of 
$
|\Omega_\kappa| \la 0.1.$
Since the horizon radius, using proper distance $d$, is 
$
R_H \approx 10{\hGpc},$
the present estimates of curvature can be succintly restated as
\begin{equation}
R_C \ga R_H \approx 10{\hGpc}.
\end{equation}

The proper distance is not usually useful in observational cosmology, 
unless phenomena in comoving coordinates are being studied. The distances
more commonly used are the {\em proper motion distance}
[p.485, \citet{Wein72}; called 
`angular size distance' by \citet{Peeb93}, p.319, eq.~(13.29)] 
\begin{eqnarray}
\dpm(z) &=&
	\left\{ 
        \begin{array}{lll}
        R_C \sinh [d(z)/R_C] , & \Omega_\kappa < 0 \\
        d(z) , & \Omega_\kappa = 0 \\
        R_C \sin [d(z)/R_C] , & \Omega_\kappa > 0.
        \end{array}
        \right.,
\label{e-defpm}
\end{eqnarray}
and distances which are greater or smaller than this by a factor of $(1+z)$.

The tangential distance of use in comoving coordinate work can be 
written as 
\begin{eqnarray}
\dperp (z,\delta\theta) &\equiv& \delta\theta \; \dpm(z) \nonumber \\
&=& \delta\theta 
	\left\{ 
        \begin{array}{lll}
        R_C \sinh [d(z)/R_C] , & \Omega_\kappa < 0 \\
        d(z) , & \Omega_\kappa = 0 \\
        R_C \sin [d(z)/R_C] , & \Omega_\kappa > 0.
        \end{array}
        \right.,
\label{e-defdperp}
\end{eqnarray}
where $\delta\theta$ is an angle in radians on the sky. 

\section{Method} \label{s-method}

Equations (\ref{e-defdprop}) and (\ref{e-defdperp}) clearly show 
that for a standard ruler placed at a large fraction of the 
curvature radius from the observer, the $\sin$ or $\sinh$ term 
will strongly distinguish the curved and flat cases. 

In order to
test for independence of orientation of a standard ruler, consider
a standard ruler 
of fixed comoving size $\llss$
which is observationally detected near a redshift $z$, radially as
a redshift interval $\delta z$ and tangentially 
as an angular size $\delta\theta$. 
Using equations (\ref{e-defdprop}) and (\ref{e-defdperp}),  
the length of the ruler placed in the radial direction is
\begin{equation}
d(z+\delta z,\Omm,\Omega_\Lambda) -d(z,\Omm,\Omega_\Lambda),
\end{equation}
and the length in the tangential direction is
\begin{equation}
\dperp(z,\delta\theta,\Omm,\Omega_\Lambda),
\end{equation} 
where the implicit dependence on the curvature parameters is now made
explicit. 
Define the difference between these as
\begin{eqnarray}
&&\Delta d(z,\Omm,\Omega_\Lambda) \equiv  \\
&&\dperp(z,\delta\theta,\Omm,\Omega_\Lambda)
- \left[d(z+\delta z,\Omm,\Omega_\Lambda) -d(z,\Omm,\Omega_\Lambda)
\nonumber
\right].
\label{e-diff}
\end{eqnarray}

Unless the Copernican principle is very surprisingly violated, 
$\Delta d$ should be equal to zero, within the measurement uncertainties
and intrinsic uncertainties of the ruler.
If the Universe is `slightly' curved, then a flat universe can be
refuted if 
\begin{equation}
\Delta d (z,\Omm,\Omega_\Lambda \equiv 1-\Omm)= 0
\label{e-theproof}
\end{equation}
can be refuted for all acceptable values of $\Omm$, or 
if the solutions to the equation are for values of $\dperp$ 
inconsistent with $L$.

The method is then defined by
\begin{list}{(\alph{enumi})}{\usecounter{enumi}}
\item choosing the acceptable range of $\Omm$, e.g. $0 \le \Omm \le 1,$
\item choosing $z$, $\llss$ and $\Delta \llss$ (the uncertainty in $\llss$),
\item calculating the implied values of $\delta z$ and $\delta \theta$,
\item assuming $\Delta \llss$ to be the total measurement and intrinsic 
uncertainty (in length units) in each of the radial and tangential
directions separately,
\end{list}
and then 
calculating
\begin{equation}
\Delta d(z,\Omm,1-\Omm) 
\end{equation}
and
the Gaussian probability that any of these values are consistent with
zero, i.e. `orientation-free', 
\begin{equation}
\Pzero= \mbox{\rm erfc}[ | \Delta d | /(\sqrt{2}\sigma) ],
\label{e-pzero}
\end{equation}
where $\sigma^2 = 2(\Delta \llss)^2$, for each value of $\Omm$.

As is seen in Fig.~\ref{f-deltad} and discussed in 
\S\ref{s-results}, this is insufficient on its own to 
rule out a flat universe hypothesis, so the additional 
hypothesis that 
\begin{equation}
\dperp(z,\delta\theta,\Omm,1-\Omm) = \llss, 
\label{e-testtwo}
\end{equation} 
is required, where the Gaussian probability is
\begin{equation}
\Pllss= \mbox{\rm erfc}[ 
| \dperp(z,\delta\theta,\Omm,1-\Omm) - \llss | /(\sqrt{2}\sigma) ],
\label{e-pllss}
\end{equation}
where $\sigma^2 = 2(\Delta \llss)^2$ as above.
(The radial ruler size could equally well be used here.)


\fdeltad

\frejecten
\frejeconep

The independence of 
the ruler from orientation and the value of the length of the ruler
are independent hypotheses, so the 
hypothesis of a flat universe is then rejected at the
\begin{equation}
 1-\Pzero \Pllss 
\label{e-pboth}
\end{equation}
confidence level.

\section{Results} \label{s-results}

Fig.~\protect\ref{f-deltad} shows that 
for reasonable values of
($\Omm,\Omega_\Lambda$) for a non-flat universe, the $\sin$ or
$\sinh$ factor which relates the radial and tangential directions,
making a ruler of different sizes if the universe 
is curved [eq.~\ref{e-defdperp}],
can be compensated for by the non-linearity in 
the distance redshift relation. In other words, even if the Universe
really is `slightly' curved, a wrong pair of values 
($\Omm,\Omega_\Lambda$), where $\Omm$ is larger (smaller) than
the true value for a hyperbolic (spherical) universe,
can be found for which the radial and tangential sizes of the standard
ruler are equal. These solutions are represented by the zero crossings
of the curves which increase with $\Omm$ in 
Fig.~\protect\ref{f-deltad}. 
The differences between the `true' $\Omm$ values and those 
required for independence of $\llss$ from orientation (for a wrong, 
flat solution) 
are not large.

This is why the size of the ruler needs to be known [eq.~(\ref{e-testtwo})]. 
The curves of $\dperp-\llss$ 
in Fig.~\protect\ref{f-deltad} (which decrease with 
increasing $\Omm$) 
are reasonably steep near both the input
and the `orientation-free, flat' 
values of $\Omm$, but to distinguish the
point of intersection of the curves 
from zero would require uncertainties of much less than
10{\hMpc}, or $\ll 10\%$ of the ruler size. 

\treject

\trejecttwo

The probabilities for rejecting the flat universe based on the
individual requirements of orientation independence 
[eq.~(\protect\ref{e-pzero})], correctness of the value of
$\llss$ [(eq.~\protect\ref{e-pllss})] and the combined probabilities
[eq.~(\protect\ref{e-pboth})] are shown in 
Figs~\protect\ref{f-rejecten} and ~\protect\ref{f-rejeconep}.
In the latter figure, higher precision, i.e. 1\%, is assumed in $\llss$.
The minimum values of $1-\Pzero\Pllss$ are listed in 
Table~\ref{t-reject}. 

For an uncertainty of $\sim10\%$, none of the `slightly' curved models, 
each having $|\Omega_\kappa|=0.1$ and $R_C \approx R_H\approx 10${\hGpc},
would enable significant rejection of a flat universe. However, 
{\em for an uncertainty of $1\%$, all three of the same 
`slightly' curved models, using data at $z=3$, would enable 
significant rejection 
of a flat universe, i.e. at the $1-P \ga 95\%$ level.}
Lower redshift data ($z=2$) 
only provides marginally significant rejection.

{
For the potentially most interesting case of $\Delta \llss = 1\% \llss$
and $z=3$, Table~\ref{t-rejecttwo} shows how the strength of the
rejection varies with $\Omm$ and $\Omega_\kappa$. A nearly flat universe
with a value of the matter density 
considerably higher than presently estimated, i.e. $\Omm \ge 0.5$,  
would be more difficult to reject, given a fixed absolute curvature
$|\Omega_\kappa|=0.1$. On the other hand, a larger absolute 
curvature ($|\Omega_\kappa|=0.2$) 
would enable refutation of a flat universe to better than $99.7\%$ 
confidence for $\Omm \le 0.5$.
}

\section{Discussion and Conclusions} \label{s-conclu}

Does a standard ruler of the required precision exist? This depends,
of course, on what the curvature of the Universe really is. If the
Universe is `exactly' flat in a geometrical sense, i.e. if the covering
space is $R^3$, then use of 
local geometrical techniques would not be sufficient to prove that
the Universe is not curved. Proof that the Universe is flat and multiply 
connected (e.g. see \citealt{LR99} for a review) would be one
way of using {\em global} geometry 
to prove that the Universe is not (on average) curved. 

However, if the radius of curvature is no bigger than the horizon, 
then the calculations above show that for reasonable values of 
($\Omm,\Omega_\Lambda$), a precision of 1\% in the application of 
a standard ruler at a redshift of $z=3$ would be sufficient. 
At higher redshifts, less precision would be needed, but the possibility
of having large surveys including sufficient amounts of 
both radial and tangential 
standard ruler information is unlikely in the next decade at $z\gg3$.

At $z=3,$ surveys of quasars or of Lyman break galaxies
(e.g. \citealt{Adel98,GiavDCP98}) of sufficient quality to detect
comoving features at large scales in the power spectrum 
should be feasible.

Whether or not fixed, comoving features in the power spectrum 
of density perturbations, as traced by these objects, exist and
are detectable,  
is still a controversial subject.
Observations by several different groups suggest that 
a peak near the maximum in the power spectrum, 
at $\llss\approx130\pm10${\hMpc}, is common to 
galaxies and superclusters of galaxies at low redshift
(\citealt{Bro90,BJ99,daCosta93,BauE93,GazB98}; 
\citealt{Einasto94,Einasto97nat,Deng96,Guzzo99,LCRS98})
and
Lyman break objects \citep{BJ99}
and quasars \citep{Deng94,RM00a,RM00b} at high redshift. 
Since $130${\hMpc} is above
the present turnaround scale, 
it should be fixed in comoving coordinates.
Moreover, several {possible}
theoretical explanations for this feature, which would
also imply other `oscillations' in the power spectrum, 
include
acoustic oscillations in the baryon-photon fluid before last scattering, 
in high baryon density models \citep{Eisen98a,MeikWP98,Peeb99b}, 
and features from sub-Planck length physics which survive through
to oscillations 
in the post-inflation power spectrum, for weakly coupled 
scalar field driven inflationary models \citep{MB00a,MB00b}. 

No group yet claims that the precision of the peak is better than 
$10\%$. Observational improvements (homogeneity of surveys, 
numbers of objects), refinements in statistical analysis techniques, and
an unambiguous theoretical explanation for the peak might help 
reduce the uncertainty in the value.

\tredacc

{
What prospects for observational improvements from surveys expected to
be completed within 1-5 years might potentially approach 
1\% precision in $\llss$?

The best possible improvements in precision would presumably scale as
Poisson errors, so that improving from $\sim 10\%$ precision to $1\%$
would require a factor of 100 increase in the numbers of objects
relative to previous surveys. 

The continuation of the original observations by \citet{Bro90}
which confirm the original result include more than 1000 
spectroscopic redshifts up to maximum redshifts of 
$z\sim0.4$ \citep{BJ99,Bro99}. The 
VIRMOS shallow survey \citep{LeFev01} is expected to obtain
100,000 spectroscopic redshifts of galaxies 
in fields of size $\sim 2\degr$ 
to a limiting magnitude of
$I_{AB}=22.5$, i.e. with typical median (maximum) 
redshifts of $z \sim 0.6$ $(z\sim 1.3)$
\citep{CFRSV}. This is probably the best near future survey which
more or less matches the conditions of the 
\citealt{Bro90} observations.
As long as the fact that the galaxies will be
spread over a somewhat larger redshift range does not adversely
affect improvements in the estimation of $\llss$, 
the factor of $\sim 100$ increase in numbers should be
sufficient to provide the precision required.

However, if the effect is anisotropic as suggested by
\citet{Einasto97nat}, then the VIRMOS shallow survey may not
be enough, since none of the four fields are close to the
directions of the \citeauthor{Bro90} fields. 

Wide angle surveys may therefore be more useful.

In the 2dF Galaxy Redshift Survey (2dFGRS, \citealt{Coll01}) it is
expected to observe spectroscopic redshifts of 250,000 galaxies over
2000 sq.deg. with a mean redshift of $z=0.1$.  For the Sloan Digital
Sky Survey (SDSS, \citealt{SDSS}), it is planned to observe about
1,000,000 galaxy spectroscopic redshifts over 10,000 sq.deg.  with a
median redshift of $z=0.1$.

Both surveys clearly provide the increase in numbers of objects
required. A possible problem in precision is the fact that
only about 100 galaxies/sq.deg. will be observed in
these surveys as opposed to around 1000 galaxies/sq.deg. in the
\citealt{Bro90} fields. If individual structures (`walls', `filaments')
are less sharply traced in position relative to the \citealt{Bro90}
surveys, then this could provide an additional noise factor.

Will spectroscopic redshift accuracy 
be sufficient {\em not} to provide an additional source of uncertainty
in the radial direction?
Table~\ref{t-redacc} shows that a precision  in redshift of
$\Delta z \ltapprox 5\e{-4}$
corresponds to $\Delta \llss=1.3${\hMpc}.

The VIRMOS survey \citep{LeFev01} is only expected to achieve
spectral resolution of $\lambda/\Delta \lambda \sim 250$ for
the full 100,000 galaxies, implying
$\Delta z \sim 4\e{-3}(1+z) \ltapprox 10^{-2}$. 
Although a sub-sample will be observed at  $\lambda/\Delta \lambda 
\sim 2500-5000$ in order to study biases, it is hard to see how 
this can be sufficient for the purposes of obtaining a precise value
of $\llss$. It would be preferable if the full 100,000 galaxy sample
could be observed at the higher resolution.

On the other hand, the spectral resolutions of the 2dFGRS and
SDSS are expected to be 
$\lambda/\Delta \lambda \sim 1000-2000$, so will provide approximately
the precision required.

However, $\Delta z \ltapprox 5\e{-4}$ precision in redshift
corresponds to $\Delta zc \ltapprox 300$km/s.  Typical galaxy velocity
dispersions in loose groups and clusters and typical bulk velocities,
or in other words, typical galaxy velocities with respect to the
comoving reference frame, are typically at about this scale or up to
nearly an order of magnitude higher.  So, averaging over these
`random' errors will be required in order 
to obtain the precision required.

In the analysis of the low redshift surveys, it should be kept in mind
that even though the redshifts are smaller than unity, a clear
and precise detection of the $\llss$ scale would require approximately
correct values of the metric parameters, and may be missed if wrong
values are used. 
}

{ 
It should also be noted that, in the hypothesis that the proposed
ruler actually exists, and can be traced back to the primordial
Universe, the theoretical explanation of the scale might heavily rely
on some assumption on yet unknown or difficult to measure cosmological
or fundamental physical parameters. This would introduce an 
additional uncertainty if the theoretical model were to be used
to define the size of the standard ruler.
}

In conclusion, if the Universe is indeed `slightly' curved, then this
method 
could potentially be
promising for proving that the Universe is not
flat.

Note that small scale clustering on quasi-linear or
non-linear scales (cf. \citealt{AP79}) would be more difficult to
use as a standard ruler, due to evolution in the ruler length.


\begin{acknowledgements}
The author thanks Gary Mamon, St\'ephane Colombi, Tarun Deep Saini,  
Varun Sahni and Thanu Padmanabhan,
 whose comments were useful and encouraging. 
The support of the Institut d'Astrophysique de Paris, CNRS, 
and of DARC, Observatoire de Paris-Meudon, 
for visits during which part of this work was carried out, 
and the support of la Soci\'et\'e de Secours des Amis des Sciences 
are gratefully acknowledged. 
\end{acknowledgements}

\end{document}